\documentclass[]{mn2e}
\usepackage{graphicx}

\def\bbb{}

\title[Light curve evolution of Kepler-13]{Mapping a star with transits: orbit precession effects in the Kepler-13 system}

\author[Gy. M. Szab\'o et al.]{
Gy. M. Szab\'o$^{1,2,3}$,
A. Simon$^{2,3}$,
L. L. Kiss$^{2,3,4}$\\
$^1$ELTE Gothard Astrophysical Observatory, H-9704
Szombathely, Szent Imre herceg \'ut 112, Hungary
\\
$^2$Konkoly Observatory, Research Centre of Astronomy and Earth Sciences, Hungarian Academy of Sciences, H-1121 Budapest,\\
Konkoly Th. M. \'ut 15--17, Hungary
\\
$^3$Gothard-Lend\"ulet Research Team, H-9704
Szombathely, Szent Imre herceg \'ut 112, Hungary
\\
$^4$Sydney Institute for Astronomy, School of Physics, University of Sydney, NSW 2006, Australia
}
\begin{document}

\date{Accepted Received; in original form}

\pagerange{\pageref{firstpage}--\pageref{lastpage}} \pubyear{2010}

\maketitle

\label{firstpage}

\begin{abstract}
 {Kepler-13b (KOI-13.01) is a most intriguing exoplanet system due to the rapid precession rate, exhibiting several exotic phenomena. We analyzed $Kepler$ Short Cadence data up to Quarter 14, with a total time-span of 928 days, to reveal changes in transit duration, depth, asymmetry, and identify the possible signals of stellar rotation and low-level activity.} 
{We investigated long-term variations of transit light curves, testing for duration, peak depth and asymmetry. We also performed cluster analysis on $Kepler$ quarters. We computed the autocorrelation function of the out-of-transit light variations.} 
{Transit duration, peak depth, and asymmetry evolve slowly, due to the slowly drifting transit path through the stellar disk. The detected transit shapes will map the stellar surface on the time scale of decades. We found a very significant clustering pattern with 3-orbit period. Its source is very probably the rotating stellar surface, in the 5:3 spin-orbit resonance reported in a previous study. The autocorrelation function of the out-of-transit light variations, filtered to 25.4 hours and harmonics, shows slow variations and a peak around 300--360 day period, which could be related to the activity cycle of the host star.}
\end{abstract}


\section{Introduction}

Kepler-13 (formerly known as KOI-13, Shporer et al. 2011, Johnson and Cochran, 2013) is a spectacular astrophysical laboratory of close-in companions in an oblique orbit. A planet-sized companion with 1.76 days period was announced as one of the 1235 Kepler planet candidates in 2011 February 
(Borucki et al. 2011, BO11 hereafter). The planetary nature of the companion was confirmed by Mazeh et al. (2012) and Mislis \&{} Hodgkin (2012). The transit curves show significant distortion that is stable in shape, and the transit curve asymmetry is consistent with a companion orbiting a rapidly rotating star on an oblique orbit (Barnes 2009, Barnes et al. 2011; Szab\'o et al. 2011, Johnson and Cochran 2013). This picture has also been consistent with the A spectral type and $v\sin l \approx 65$ $km \ s-1$ {\bbb rotation velocity ($l$ is the inclination of the stellar spin).}

\begin{figure}
\includegraphics[width=\columnwidth]{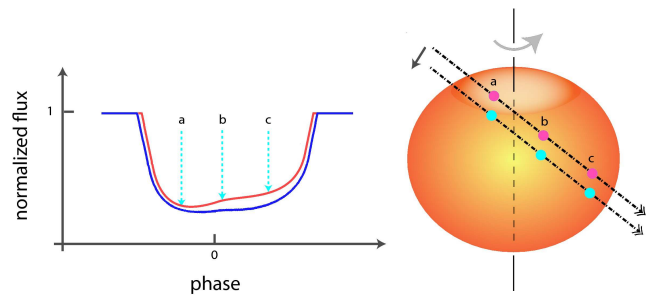}
\caption{Illustration of how a transiting planet on a precessing orbit can map the entire stellar surface. Left panel shows the expected differences in light variation.}
\end{figure}

Variation of transit duration was first reported by Szab\'o et al. (2012) for Q2--Q3 data and was confirmed by Mazeh et al. (2012) for Q2--Q9. This effect is due to the precession of the orbital plane around the total angular momentum vector of the system, which causes a decrease in the impact parameter $b$, and the transit path gets closer to the center to the stellar disk. After several decades, the orbit will turn over and even will leave the visible disk of the host star, and Kepler-13 will thus become a non-transiting system (Szab\'o et al. 2012). During this process, the stellar surface will be mapped by the transit light curve, offering a unique opportunity to reveal the brightness distribution of an exoplanet host star with high precision and resolution (Fig. 1).

\begin{figure*}
\includegraphics[bb=1 1 841 490,width=2.0\columnwidth]{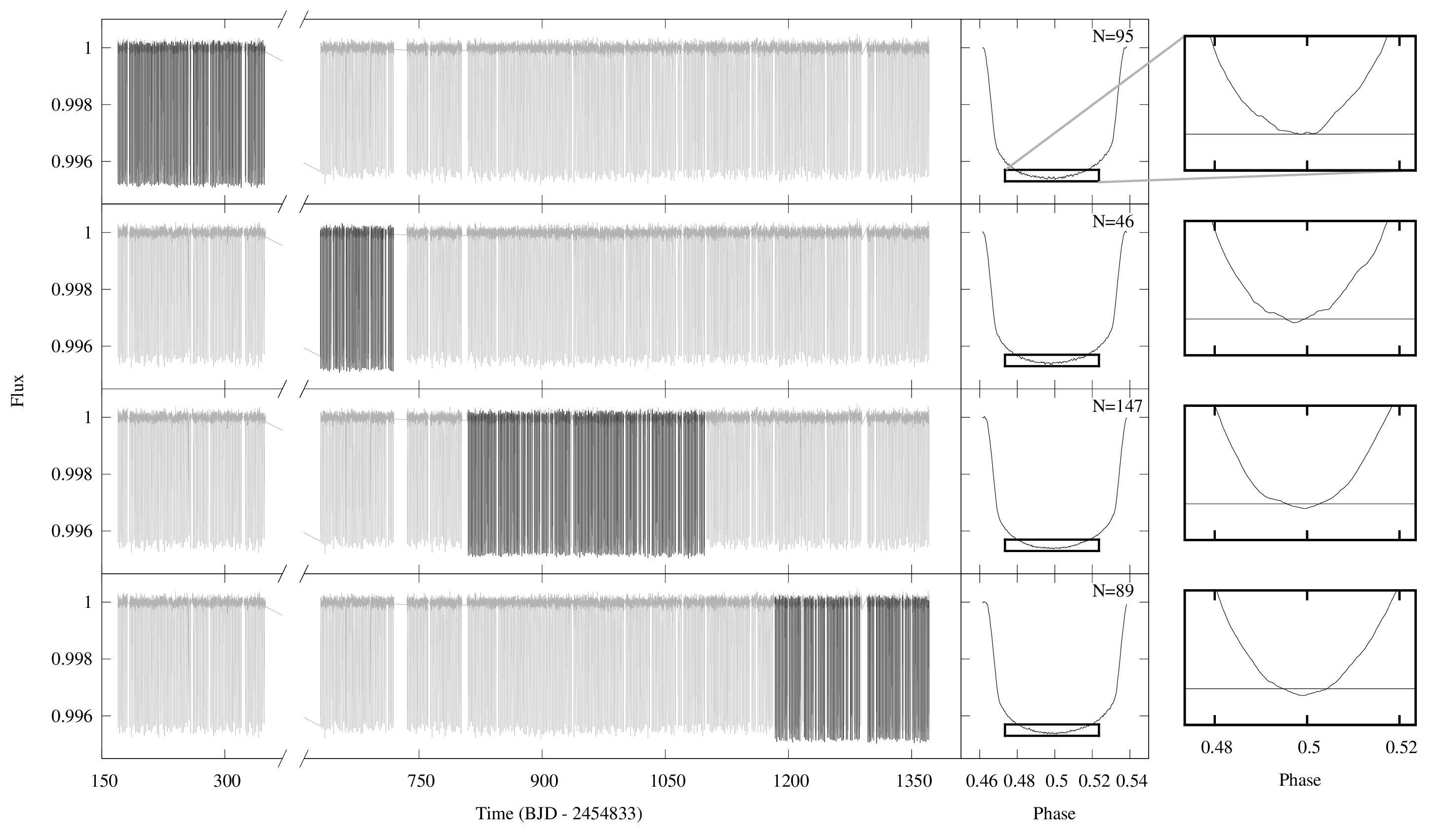}
\caption{Long-term variation of transit curve peak depth, due to orbit precession. Left panels: the analyzed segments are shown by black overplots on the full $Kepler$ lightcurve. Middle panels: the average light curve shape, calculated in the consecutive segments. Right panels: a magnification of the average light curve shape, showing the minimum of transit. Deepening light curve shapes are compared to the same, constant flux level in each segments (see the straight line).}
\end{figure*}

Beside the light variation related to transits, a second periodic signal was discovered with 25.4 hour period. There is a controversy about the origin of this signal, that could either be pulsation (Mazeh et al. 2012, Shporer et al. 2011) or stellar rotation and activity (Szab\'o et al. 2012). In the latter case, the companion would orbit in 3:5 resonance to stellar rotation, on a special orbit which is quasisynchronous at large siderocentric latitudes. Similar spin-orbit resonances are known for few other exoplanet systems (e.g. KOI-63, Sanchis Ojeda 2011, Kepler-17, D\'esert et al. 2011), however the dynamical origin of this resonance is debated. Arguments for the rotational origin of the 25.4 hour signal is therefore very important, since Kepler-13 is suitable for a detailed and precise dynamical modeling, and may act as the key in understanding spin-orbit resonances of hot Jupiters in the future.

In this paper, we report on the first observation of the visible changes in transit light curves. We examine the transit depth and the asymmetry of the light curve, and compare their evolution to Barnes (2009) models  in Sect. 2. We point out that the well-known 25.4-hour period signal (Mazeh et al. 2011, Szab\'o et al. 2012, Shporer et al. 2011) also shows a time evolution with a suspected correlation timescale of 300--360 days. We discuss its possible origin in Sect. 3.

\section{Long-term variations of transit light curve shape}

The discovery of precession, and the shifting of transit path toward lower latitudes (Szab\'o et al. 2012) predicted secular variations in light curve shape, such as its depth and asymmetry. To illustrate the features, we show average light curve shapes calculated in four different segments of the entire dataset in Fig 2. Since the light curve itself is quite complex, specific observables need to be derived to quantitatively describe the subtle long-term variations. We introduce these quantities in the followings.

\subsection{Variation of $q90$ light curve depth and asymmetry}

The changing transit duration  and the prominent distortion in the light curve makes difficult to characterize the transit depth using the standard symmetrical models that are unable to describe the asymmetries. To evaluate the observation, we defined two non-parametric quantities which characterize the light curve depth and its asymmetry in a model-independent and robust manner. The $q90$ depth of the transit light curve is defined as the 90\%{} percentile of occulted light in the pre-processed (Sect. 2.2) transit curves. Practically this separates the lowest 10\%{} of transit points from the higher parts. 

The light curve asymmetry was expressed by the phase lag between the median time of the bottom 10\%{} of the light curve, and between that of the entire transit shape (more precisely, the part of the transit which exhibits more occulted light than 10\%{} of $q90$). This quantity is defined in the phase domain instead of time, because we suggest its determination for several consecutive transits which will be folded. This is a very beneficial approach, because estimating $q90$ and its median time is noisy in case of Kepler-13 due to low number statistics (one transit light curve contains $\approx$220 points). So the transit asymmetry was expressed as
\begin{eqnarray*}
\phi_{q90}-\phi_{q90/10} = \rm{Median} (\phi_i \ \ | \Delta m_i < q90) - \\ 
- \rm{Median} (\phi_j \ \ | \Delta m_j < q90/10),
\end{eqnarray*}
\begin{equation}
\end{equation}
where $\phi_i$ is the phase of the $i$-th photometric point involved in this analysis, $\Delta m$ is the occulted light (one can use either flux or magnitudes in this formalism, due to their monotonic relation and the non-parametric nature of $\phi_{q90}$).

In Fig 3., we plot the secular variation of three parameters. The top panel shows the increment of transit duration up to Q13. The linear fit is still satisfactory. The middle and bottom panels show the variation in $q90$ transit depth and in transit asymmetry. The light curve was divided into four observation sections, while we omitted Q7 and Q11 because of instrumental instabilities after safe mode gaps (transit depth is anomalously large and unstable in the omitted regions). Small dots represent individual transits for the transits involved in our analysis; while the large dots with error bars are derived from folded light curves for each sections. 

Both transit depth and asymmetry increases significantly. Q90 transit depth has increased from 4600 to 4620 ppm (uncorrected for the light from KOI-13 B), which means a $\approx$ 5 percent variation, and is very significant. The explanation is that during the migration of the transit path, the brightest part of the occulted cord has replaced closer to the center of the stellar disk, which exhibits about 5\%{} higher surface brightness. This observations is a second evidence for the precession of the orbital plane, independently from the already registered transit duration variations. 

The evolution of transit asymmetry is in an apparent contradiction with the light curve model of Barnes et al. (2011). In this model, due to the decreasing mean latitude, transit path moves away from the hot spot, and the asymmetry due to the hot spot should relax with time. {\bbb The confidence of the decreasing trend is, however, not too high (somewhat below 95\%{} according to a test of Pearson's correlation coefficient), therefore a convincing conclusion would require more observations.}

\begin{figure}
\begin{center}
\includegraphics[bb=234 184 380 560,height=\columnwidth,angle=270,clip]{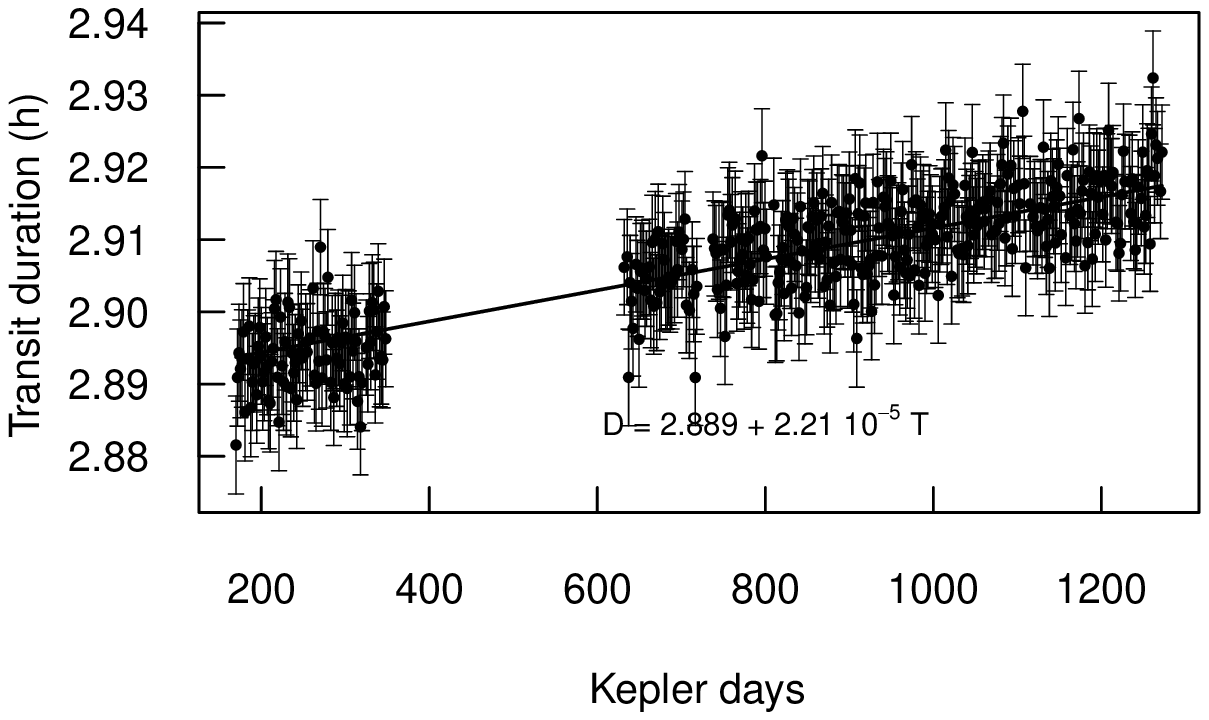}
\includegraphics[bb=234 184 380 560,clip,height=\columnwidth,angle=270]{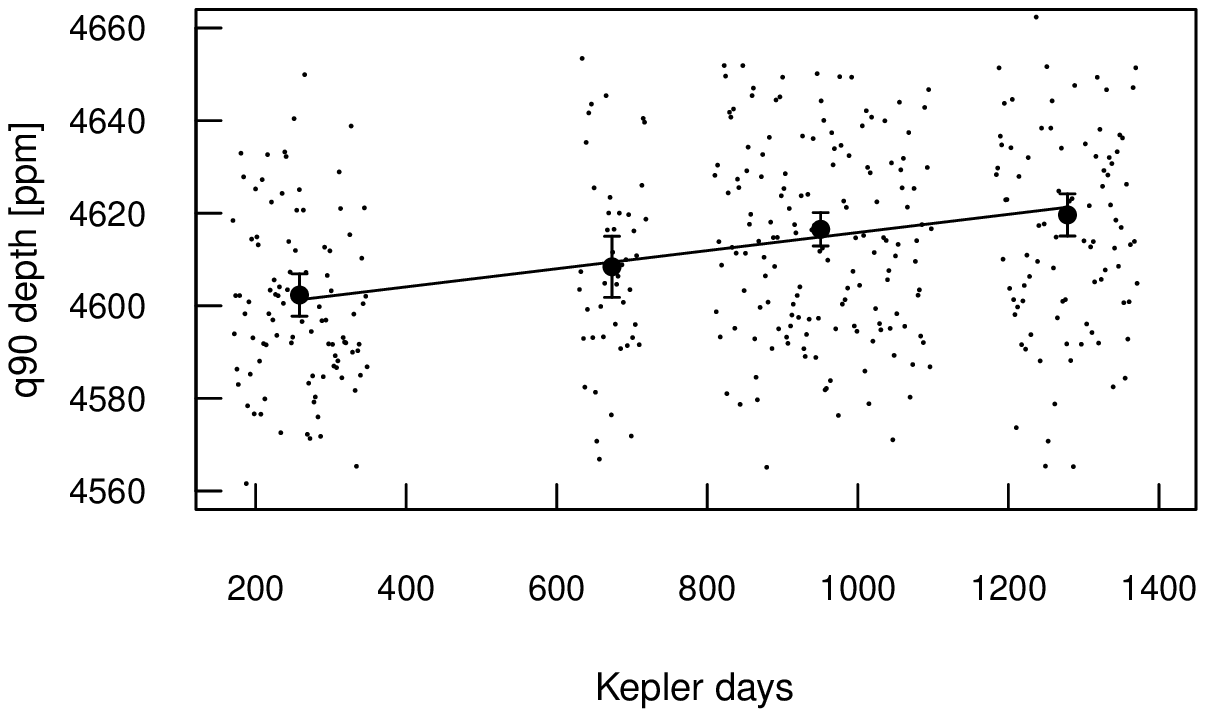}
\includegraphics[bb=234 184 450 560,height=\columnwidth,angle=270]{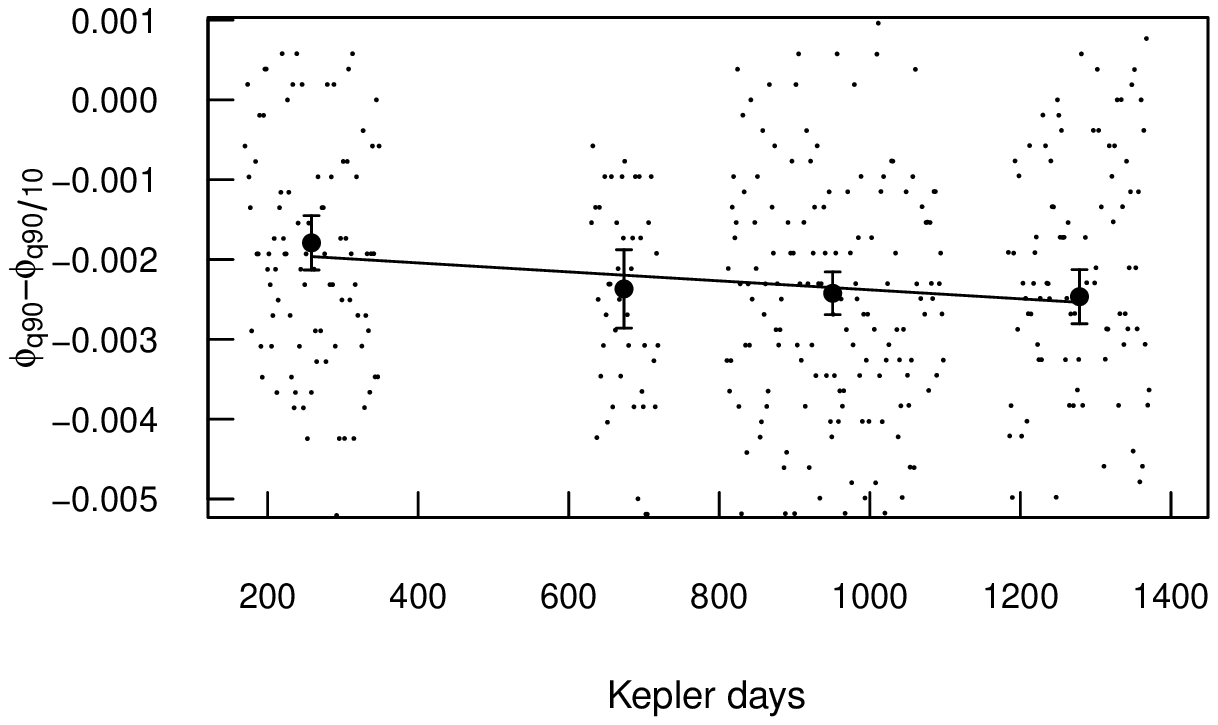}
\caption{Upper panel: Variation of transit duration of Kepler-13. Middle panel: Variation of the transit depth, expressed in term of $q90$. Lower panel: variation of the light curve asymmetry, characterised by $\phi_{q90}-\phi_{q90/10}$.}
\end{center}
\end{figure}

\subsection{Clustering analysis as argument for spin-orbit resonance}

\begin{figure}
\includegraphics[bb=50 323 522 476,width=1\columnwidth]{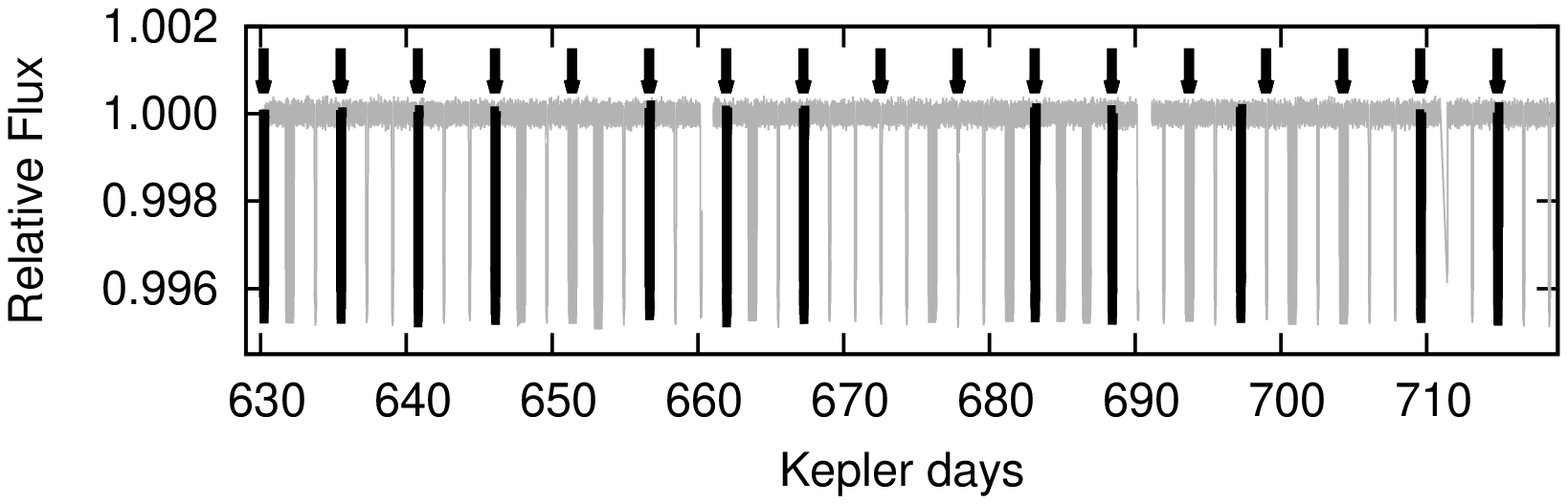}

~~~~~~~~~~~~~\includegraphics[bb=243 251 377 540,height=0.85\columnwidth,angle=270]{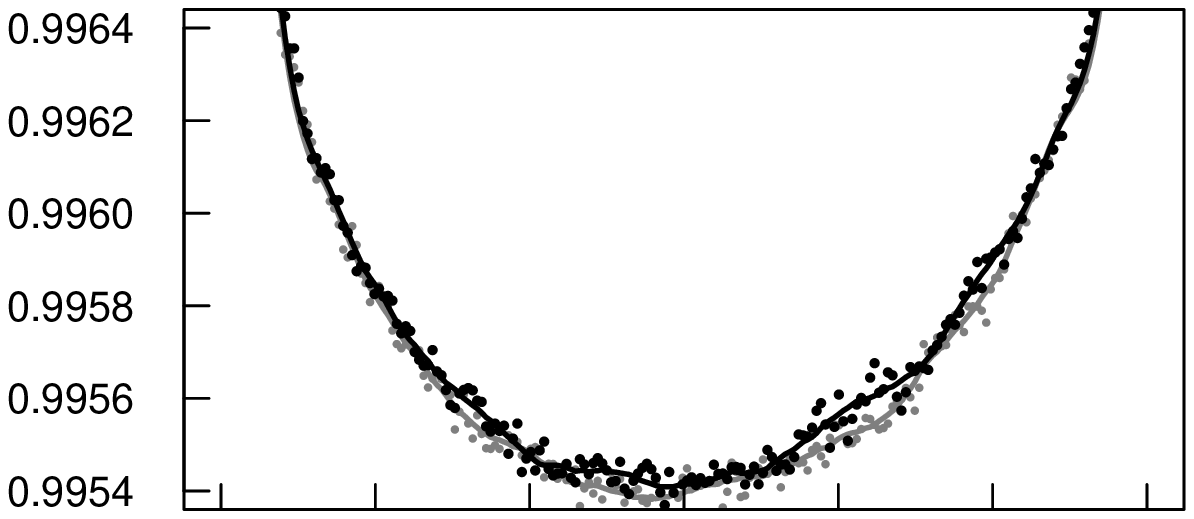}\vskip0.5cm

~~~~~~~~~~~~~\includegraphics[bb=270 252 400 539,height=0.848\columnwidth,angle=270]{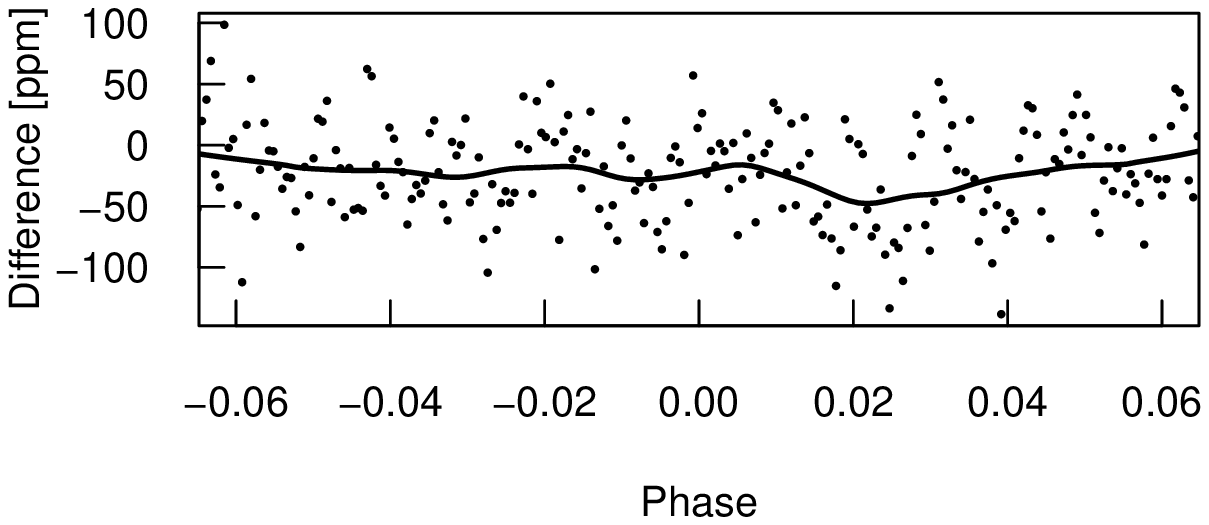}
\caption{Clustering of transit shapes. Upper panel: clusters defined by shape return in a triplet period pattern. Middle panel: Comparison of light curves in ``black'' and ``grey'' clusters, according to coloring in the upper panel reveal surface brightness fluctuations. Lower panel: the difference between ``black'' and ``grey'' clusters.}
\end{figure}

To reveal the internal structure due to e.g. repeating transit curve anomalies, we applied a hierarchical clustering (e.g. Zappal\'a et al. 1995 for an application) to the light curve shapes in 90-day long sections (where stellar surface is thought to be self-consistent enough to reveal any rotation effects). 

Pre-processing included the de-stretching of the time coordinate of individual light curves to correct for increasing transit duration (see Fig. 3 for the applied linear trend taken into account). Systematics were removed with a slowly changing filter (smoothly balanced cubic spline model fitted to the out-of-transit light variations), and the light curve was normalized to unit out-of-transit flux. Point defects (single outliers) were identified in the folded light curve, and they were replaced by a model flux interpolated from the neighboring data points in the folded light curve. Finally, light curves were de-folded into time domain, and all transits were resampled equidistantly and densely at a common phase comb for all transits. 

The distance matrix of light curves was characterized by their Euclydian distance in terms of occulted light,
\begin{equation}
D_{k,l} = g_{i,j}\ \left( {  \sum_i {(\Delta m_{k,i}-\Delta m_{l,i})^2 \over (err_{l,i}^2 + err{k,i}^2)  } } \right) ^{1/2},
\end{equation}
where $\Delta m_{j,i}$ is the occulted light in the $i$-th light curve point of the $j$-th transit normalized to the mean out-of-transit flux., and $err_{j,i}$ is its error in the $Kepler$ photometry data table. The statistical weights (scaling the internal scatter of the light curve properly) were set to $g_{i,j}\equiv 1$ initially. After identifying the closest neighbors in the matrix, these two transit curves were averaged together, and the new, $(N-1)\times (N-1)$ distance matrix was calculated. During the evolution of clustering, weights were set to $g_{i,j}=[nm/(n+m)]^{1/2}$, where $n$ and $m$ counted the original light curves averaged into the $i$th and $j$th pattern.\footnote{In case of stationary noise distribution with $\nu$ variance, the variance of noise in the $i$th transit curve is $\nu/n$;the same for $j$; and if the two transits are equivalent in shape, their squared distance will be proportional to $\nu/n + \nu/m$. The calculation of the statistical weight is straightforward from here.} The tree was cut at a three branches state (i.e., after reducing the data set to three light curves, we interpreted these light curves as cluster centers showing ``typical light curve shapes'', and the concatenating process also revealed the clustering vector, i.e. which transits belong to which cluster). 

The clustering pattern clearly shows a repetitive nature, suggesting that every third transits are similar in shape. We plot the results for Q6 in Fig. 4. The upper panel shows the clustering vector coded by colours. Note that black-coded transits, that we call the ``black'' cluster in the followings, show a clear pattern with 3-cycle period. The exact pattern is illustrated by the arrows. Besides 11 coincidences between the clustering vector and the arrows (true positive), there is only one false positive (a cluster member without arrow). Moreover, 6 false negatives (arrow without membership) and 27 true negatives (no arrow, not member) are observed. This contingency is statistically very significant, a similar periodic pattern occurs with only 1:2000 probability from random fluctuations as calculated by a Pearson's Chi Square Test for independence of categorical variables (Sokal et al. 1995). The other two ``gray'' clusters are closer to each other, however, the one plotted with wide lines has also an 1:85 contingency significance. 

Although this finding for Q6 is the most significant, clustering pattern with 3-orbit period is drawn similarly for other $Kepler$ quarters. The recurrence of similar light curves with 3-orbit period is a strong evidence for that every third transit occurs in front of a similar stellar disk. That leads to the conclusion that the stellar rotation and the orbital motion of the planet is indeed in a 5:3 resonance; and the origin of the 25.4 hour period is the stellar rotation itself. 

The middle and bottom panels of Fig. 4 reveal the origin of the pattern that led to systematic differences in transit light curves, and consequently, the clustering pattern: a prominent spot-like feature is present in ``black'' cluster members around 0.02 phase. The difference curve also oscillates around a non-zero mean, ``black'' transits are shallower by 22 ppm in average. The Student-t statistics (Lupton 1993) of differences (denoted by black points in the bottom panel) is 7.9 against a constant zero assumption, assuring a very high significance for the detected differences and periodicity in transit depth.

\subsection{Evidence for rotation in the Fourier spectrum}

\begin{figure}
\begin{center}
\includegraphics[bb=58 207 460 500,width=\columnwidth]{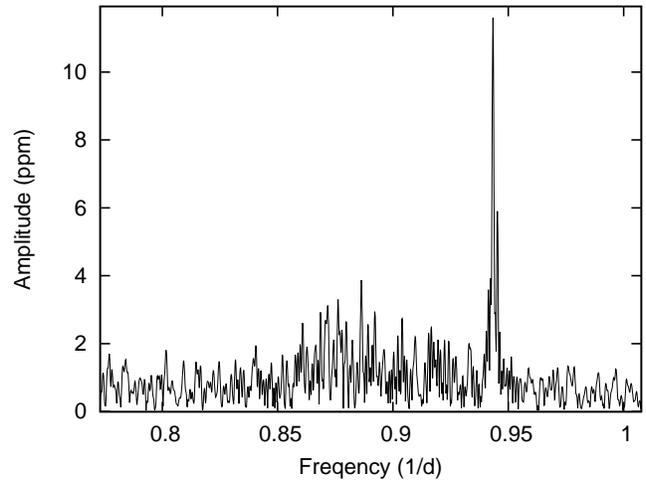}
\caption{Fourier spectrum of Kepler-13 light curve near the 25.4 hour peak. The ``common feature'' of rotating A-stars (Balona 2013) can be seen at the left side of the peak.}
\end{center}
\end{figure}

In a series of studies on activity in A-type stars, Balona (2011, 2012, 2013) reported signs of activity in 40\%{} of A-type $Kepler$ targets, detectable during the typically 800-day long observation window. He concluded that A-type stars are similarly active as Sun-like stars, with similar spot sizes and differential rotation as on the Sun. However, the equatorial velocities of active A-type stars are distributed identically to all A-type stars, i.e. they are rapid rotators contrary to the solar-like stars.

The rotation is accompanied with characteristic spectral features. One of them is called a ``common feature'' (Balona 2013), which is a forest of peaks under a wide (typically extended to $\approx$0.05 day) envelope, at the low-frequency side of the narrow peak belonging to the rotational frequency. {\bbb There is no solar analogy for such a peak, which yet has an unknown origin. However, is clearly observed in the majority of rotating A-type stars.}

Another interesting characteristics of active A-type stars is a splitting of the rotation peak (Balona 2013). The multiple structure has been interpreted as a sign of differential rotation. Once the active latitudes exhibit different rotation velocities, these will emerge in the form of separate peaks in the spectrum, numbering 4--5 in some cases.

In the case of Kepler-13, the spectrum of the out-of-transit signal does show (Fig. 5) both spectral features detected by Balona (2013). The ``common feature'' is apparent, and extends to 0.85 cycle/day on the long-period side of the rotational peak. The peak is multiple, exhibiting two significant components at 0.9433 and 0.9453 1/day frequencies (25.389 and 25.426 hour periods). These findings are compatible with the picture that the host star of Kepler-13 indeed rotates with 25.4 hour period. The period itself is also compatible with rotation: A-type stars in $Kepler$ field exhibit a complex period distribution, consisting a gradient toward longer periods, and a significant hump superposed around 1 day. The 25.4 hour period observed in Kepler-13 is exactly in the middle of the hump.

\section{Long-term behaviour of the 25.4-hour signal}

\begin{figure}
\begin{center}
\includegraphics[bb=118 241 546 510,height=\columnwidth,angle=270]{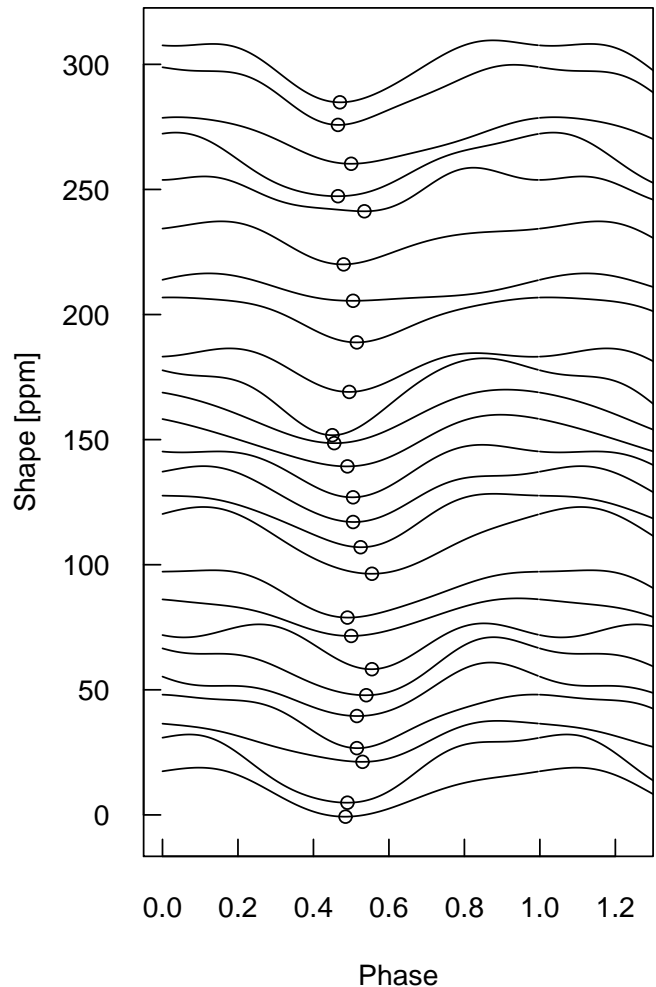}
\caption{Variation of the out-of-transit light curves. Phase diagrams with 25.4 h period are plotted for consecutive 30-day long observation windows. Global minima are also indicated, to emphasize the relative similarity of neighboring segments.}
\end{center}
\end{figure}

The out-of-transit light variation was found to exhibit at least two components: there is light variation related to the orbit of Kepler-13.01 (ellipsoidal variation, beaming, reflected light), and there is another component with 25.4 hours period of disputed origin. Mazeh et al. (2012) and Shporer et al. (2011) {\bbb argue} for its pulsation origin, while our group (Szab\'o et al. 2012) collected several arguments for a possible activity origin, hence the rotation period of Kepler-13 should be 25.4 hours, in 3:5 resonance with the orbital period.

Here we intend to reveal slow, time-dependent evolution of the 25.4-hour signal, with a possible correlation length of 300--360 days. Then we compare this result to the expectations from the most recent studies on activity among A-type stars.

Before the analysis for internal correlations, light curves were pre-processed according to the following algorithm:
\begin{enumerate}

\item{} {\bbb The SC light curve} was segmented into $\approx$30-day long sections, following the natural segmentation of $fits$ files of $Kepler$~SC observations (where each quarters are segmented into 3 parts);

\item{} Transits and secondary transits were masked out;

\item{} Systematics were removed and the light curve was normalized to a 3rd order polynomial fitted to each segments;

\item{} With Fourier-filtering, we removed all variations related to the orbital period and its harmonics (i.e. applying narrow period cuts with the orbital period and up to its 6th harmonics);

\item{} After re-transforming back to the time domain, we produced folded light curves with 25.4 hour period;

\item{} Which were resampled densely and equidistantly, and were sent to the correlation test.

\end{enumerate}

\begin{figure}
\begin{center}
\includegraphics[bb=188 207 472 550,height=\columnwidth,angle=270]{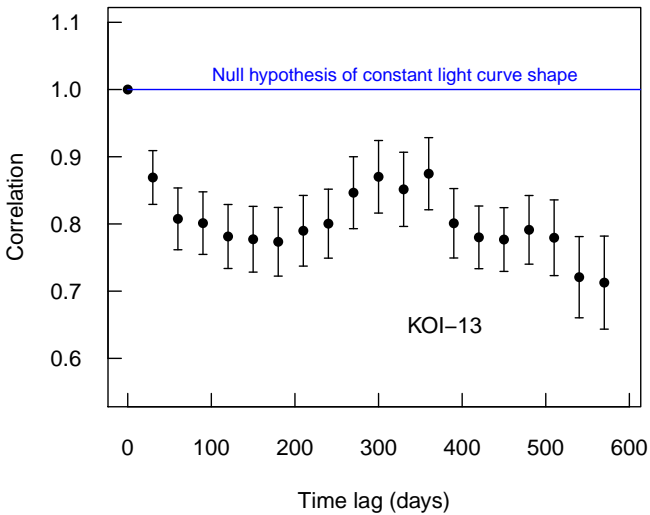}
\caption{Cross-correlation function of out-of-transit light curve shapes. The continuous evolution is evident. Note the secondary peak at about 300--360 day period.}
\end{center}
\end{figure}

This way, we got 30 individual folded light curves, containing the mean shape related to the 25.4-hour variations, at $\approx$ 30 days distance between the neighbors (there was a gap because no SC data were taken in Q3-4). These light curves are plotted in Fig. 6. The recurrent behavior of the light variation can be recognized by eye. To demonstrate the recurrence of an important light curve feature, we marked the minimum of each light curves, {\bbb which varies according to a random walk, i.e. it ``has memory''. If there would be noise only due to uncorrelated numerical fluctuations by photometry errors, we would see uncorrelated fluctuations around a stable position.}

The correlation of light curves at $k$ time lag was defined as
\begin{equation}
C(k)=\sum_j { \sum_i [ m_{j,i} \times m_{j+k,i} ] \over [N-k]},
\end{equation}
where $m_{x,i} \ \ (x=j; j+k)$ is the stellar brightness belonging to the $x$th (filtered and resampled) light curve at $\phi_i$ phase, $k$ is the time lag, and $j$ is a running index between 1 and $N-k$ ($N$ is the total number of light curves, therefore $1<j+k<N$). We emphasize that the cross-correlation is calculated for all light curves that are $k$ time apart, and then the mean is calculated. Then $C(k)$ was normalized to the value measured in the autocorrelation case,
\begin{equation}
{\rm Correlation} = {C(k) \over C(k\equiv 0) }.
\end{equation}

The error bar of each correlation points was calculated with bootstrapping: we resampled 1000 bootstraps of the 30 folded light curves, and calculated the correlation value of these samples where future and past are independent (i.e. they have a ``memoryless'' property). As it was expected, the bootstrap samples oscillated around 1.0 correlation values (representing that the simulations are  compatible with a stable process and superposition of some uncorrelated stochastic noise); and the error bar at $k$ time lag was calculated as the standard deviation of the bootstrap correlations at $k$.

The correlation function is plotted in Fig. 7. The dots represent correlation values, the error bars are boostrap standard deviations. For comparison, we plotted a line at 1.0, representing a stable light curve shape.

The observed variation is clearly incompatible with stability. Instead, the correlation decreases until $\approx$ 170 day time lag, showing that the light curve suffers systematic variations, walking away from the initial state. However, later the correlations start increasing later, and a secondary maximum is observed at $\approx$300--360 day time lag, with $\approx$0.85 correlation level. This behavior shows that the process ``has memory'', i.e. the random walk is regulated by some internal process with 330 day period, evolving the 25.4 hour period variations close to the initial state. It is interesting that the time lag we found is compatible with the long-period amplitude variations of active A-type stars, which is characteristically a some 100 day long process. Balona (2013) suggests that the most straightforward explanation can be signs of activity cycle; however, 3-year long $Kepler$ time series are too short to draw the firm conclusion.

\section{Summary}

In this paper we uncovered the following novelties about the Kepler-13 system:
\begin{itemize}

\item{} We found a systematic variation of the peak transit depth, which is related to the precession of the orbit. This has been the first detection of long-term transit depth variations. We also reported the variation of transit asymmetry.

\item{} With clustering analysis of $Kepler$ cadences, we have proven that the 25.1 hour signal in the light curves really comes from the host star, thus implies a significant star-planet interaction. Our method can be used more generally to uncover recurring surface features of exoplanet host stars.

\item{} We found proof for the rotation origin of the 25.1-hour period, and excluded the pulsation scenario; and

\item{} We found a recurring or probably quasiperiodic signal from the host star, which may be linked to the cycle of the stellar activity.

\end{itemize}

The discussed observations about the 25.4 hour variation are compatible with its suspected origin in stellar rotation and activity. The splitted peak and the ``common feature'' in the spectrum are exact analogies of Balona's findings in active A-type stars. With clustering analysis, we revealed that the transit shapes are affected by the suspected rotation period, therefore the source of the 25.4 hour signal is evidently Kepler-13, the host star of the transiting companion.\footnote{Should KOI-13~B, a bright stellar companion on the same $Kepler$ pixel be the source of the rotation-like signal, transit shapes would be unaffected - contrary to our findings).} In summary, the evidence strengthens the interpretation of a suspected spin-orbit resonance (Szab\'o et al. 2012) in Kepler-13. Considering this and other similar systems (KOI-63 for example), it may be plausible that spin-orbit resonances are dynamically preferred at least in some exoplanet systems, and they play a role in the evolution of planetary orbits or host star rotation - another unsolved question in exoplanet science.

\section*{Acknowledgments}
This project
has been supported by the Hungarian OTKA Grants K76816,
K83790, K104607, the HUMAN MB08C 81013 grant of the
MAG Zrt., the ``Lend\"ulet-2009 Young Researchers'' Program
of the Hungarian Academy of Sciences and by the City of
Szombathely under agreement No. S-11-1027. GyMSz
was supported by the J´anos Bolyai Research Scholarship of the
Hungarian Academy of Sciences.

{}

\end{document}